\documentclass[aps,prl,twocolumn,showpacs,floatfix,nobibnotes,superscriptaddress,amsmath,amssymb,longbibliography]{revtex4-1}
\usepackage{graphicx}
\usepackage{epsfig}
\usepackage{bm}
\usepackage{color}
\usepackage{float}
\usepackage{dcolumn}
\usepackage{multirow}
\usepackage{multibib}
\usepackage{hyperref}
\usepackage{mathtools}
\usepackage{xspace}

\newcites{app}{Supplemental References}

\graphicspath{{.}}

\newcommand{\be}{\begin{equation}}
\newcommand{\ee}{\end{equation}}
\newcommand{\ba}{\begin{eqnarray}}
\newcommand{\ea}{\end{eqnarray}}

\DeclareMathAlphabet{\mathpzc}{OT1}{pzc}{m}{it}

\begin{document}

\title{How chiral forces shape neutron-rich Ne and Mg nuclei}

\author{A.~Ekstr\"om}
\affiliation{Department of Physics, Chalmers University of Technology, SE-412 96 G\"oteborg, Sweden}

\author{C. Forss{\'e}n}
\affiliation{Department of Physics, Chalmers University of Technology, SE-412 96 G\"oteborg, Sweden}

\author{G.~Hagen}
\affiliation{Physics Division, Oak Ridge National Laboratory, Oak Ridge, Tennessee 37831, USA}
\affiliation{Department of Physics and Astronomy, University of Tennessee, Knoxville, Tennessee 37996, USA}

\author{G.~R. Jansen}
\affiliation{National Center for Computational Sciences, Oak Ridge National Laboratory, Oak Ridge, TN 37831, USA}
\affiliation{Physics Division, Oak Ridge National Laboratory, Oak Ridge, Tennessee 37831, USA}

\author{T.~Papenbrock}
\affiliation{Department of Physics and Astronomy, University of Tennessee, Knoxville, Tennessee 37996, USA}
\affiliation{Physics Division, Oak Ridge National Laboratory, Oak Ridge, Tennessee 37831, USA}

\author{Z.~H.~Sun}
\affiliation{Physics Division, Oak Ridge National Laboratory, Oak Ridge, Tennessee 37831, USA}

\begin{abstract} 
We compute the structure of the exotic even nuclei $^{20-34}$Ne and
$^{34-40}$Mg using interactions from chiral effective field theory
(EFT). Our results for the ground-state rotational bands in
$^{20-32}$Ne and $^{36-40}$Mg agree with data. We predict a
well-deformed $^{34}$Ne and find that $^{40}$Mg exhibits an oblate
deformed band close to the prolate ground-state, indicating the
emergence of shape co-existence at the neutron dripline. A global
sensitivity analysis shows that the subleading singlet $S$-wave
contact and a pion-nucleon coupling strongly impact deformation in
chiral EFT.
\end{abstract}

\maketitle

{\it Introduction.---} Neutron-rich nuclei beyond the magic neutron
number $N=20$ are interesting because of the breakdown of this shell
closure and the interplay between nuclear deformation and weak binding
in the so-called island of
inversion~\cite{warburton1990,terasaki1997,rodriguezguzman2003,yamagami2004,kosho2011,sumi2012,li2012,caurier2014,marevi2018}. In
neon (proton number $Z=10$) the dripline nucleus is
$^{34}$Ne~\cite{lukyanov2002,ahn2019}, and signatures of rigid
rotation (with relatively low-lying $J^\pi=2^+$ states whose energies
decreases with increasing $N$) are found for isotopes with
$N=20,22$~\cite{yanagisawa2003,doornenbal2009,doornenbal2016,murray2019}. In
magnesium ($Z=12$) shape
co-existence~\cite{heyde1991,gade2016,togashi2016} has been observed
in $^{32}$Mg~\cite{wimmer2010}, the dripline is thought to be beyond
$N=28$, and nuclei are deformed for $N\ge
20$~\cite{motobayashi1995,iwasaki2001,church2005,elekes2006,baumann2007,michimasa2014}. The
structure of the weakly bound dripline nucleus $^{40}$Mg is puzzling
and intriguing: The lowest $0^+$ and $2^+$ states are known and it is
not clear how to place a state associated with a second observed
$\gamma$-ray transition into the spectrum~\cite{crawford2019}. Recent
calculations suggest that the low-lying spectrum signals shape
co-existence, and that the coupling to continuum degrees of freedom
impacts its structure~\cite{tsunoda2020,macchiavelli2022}.

In this Letter we revisit even neon and magnesium nuclei from an
\textit{ab initio} perspective~\cite{Ekstrom:2022yea}, building on the
most recent works~\cite{takayuki2020,Frosini:2021sxj,hagen2022}, and
aiming at two goals: First, we use the chiral interaction 1.8/2.0~(EM)
of Ref.~\cite{hebeler2011} to more accurately predict the structure of
the dripline nuclei $^{34}$Ne and $^{40}$Mg; we also employ an
ensemble of chiral potentials to arrive at quantified uncertainties.
Second, we employ emulators and a global sensitivity analysis to
investigate how chiral interactions impact deformation in this region
of the nuclear chart.

The question ``What drives nuclear deformation?''  has captivated
generations of nuclear physicists. We briefly summarize relevant
milestones: \textcite{bohr1952,bohr1953,nilsson1955} explained
deformations as the surface vibrations of a liquid drop and the motion
of independent nucleons confined inside~\cite{bohr1975}.  In the
1960s, \textcite{baranger1968} performed Hartree-Fock Bogoliubov
calculations in two active shells and showed that the competition of
pairing and quadrupole interactions determine nuclear
deformation~\cite{kumar1968}. A decade later \textcite{federman1979}
demonstrated that deformation in the shell model sets in when
isoscalar neutron-proton interactions dominate over isovector pairing
interactions.  \textcite{dufour1996} revisited deformation in the
nuclear shell model and found it useful to decompose the Hamiltonian
into monopole and multipole parts~\cite{zuker1995,duflo1999}.  Here,
the monopole essentially is the one-body normal-ordered term of the
shell-model interaction, while the multipole terms are two-body
operators; they contain the residual pairing and quadrupole
interactions.  These results have been succinctly summarized by
Zuker's ``Multipole proposes, monopole disposes''\cite{poves2018},
i.e., the competition between pairing and quadrupole interactions
might suggest deformation while the monopole---the effective spherical
mean field---acts as a referee.

We clearly have a good phenomenological understanding about nuclear
deformation but lack a more microscopic understanding of which parts
of the nucleon-nucleon interaction impact deformation. While the
pairing interaction is readily identified with the nucleon-nucleon
interaction in the $^1S_0$ partial wave, the origin of the quadrupole
interaction is opaque. With view on Ref.~\cite{federman1979}, one
might be tempted to identify the quadrupole interaction with the
isoscalar $^3D_2$ partial wave (which is attractive). However, the
quadrupole interaction is long range---in contrast to the short-range
nucleon-nucleon interaction---and it is applicable only in model
spaces consisting of one-to-two shells~\cite{baranger1968}. Thus, our
understanding of nuclear deformation is still limited to a
low-resolution picture. The \textit{ab initio}
computations~\cite{caprio2015,dytrych2020,takayuki2020,Frosini:2021sxj,hagen2022,becker2023}
reproduced deformed nuclei but did not investigate how they are shaped
by the underlying forces. In this work, we seek to understand what
impacts deformation at the highest resolution scale possible today,
i.e., based on chiral effective field theory (EFT)~\cite{epelbaum2008,
  machleidt2011,Hammer:2019poc}. This is currently as close as we can
get in tying low-energy nuclear structure to quantum chromodynamics
(QCD) without actually solving QCD.

{\it Hamiltonian, methods, and model space.---} We use the intrinsic
Hamiltonian
\be H = T - T_\mathrm{CoM} + V_{NN} + V_{NNN}.
\label{Eq:hamiltonian} 
\ee Here $V_{NN}$ is the nucleon-nucleon ($NN$) potential, $V_{NNN}$
the three-nucleon ($NNN$) potential, $T$ the total kinetic energy, and
$T_{\mathrm{CoM}}$ the kinetic energy of the center of mass. We employ
the chiral $NN$ and $NNN$ interaction 1.8/2.0 (EM)~\cite{hebeler2011},
which yields accurate binding energies and spectra of light-, medium-,
and heavy-mass
nuclei~\cite{hagen2015,hagen2016b,simonis2017,morris2018,gysbers2019,hebeler2023}. We
also used an ensemble of about $100$ chiral (next-to-next-to-leading
order) NNLO $NN$ and $NNN$ interactions with explicit Delta degrees of
freedom (with non-local regulators and a momentum cutoff of
394~${\mathrm{MeV}}/c$ ~\cite{ekstrom2018,jiang2020}). This ensemble
consists of non-implausible interactions filtered out by history
matching~\cite{vernon2010,vernon2018,hu2022} to scattering phase
shifts, deuteron properties, the $^3$H, $^4$He, $^{16}$O binding
energies and charge radii, and ground- and excited states in
$^{22,24,25}$O~\cite{kondo2023}.  We assigned posterior weights
conditional on the $J^\pi=2^+$ and $4^+$ rotational states in
$^{24}$Ne and then used importance
resampling~\cite{smith:1992aa,Jiang:2022off} to make posterior
predictive distributions for rotational states in other
nuclei. \textcite{hu2022} used a similar approach to predict the
neutron-skin thickness of $^{208}$Pb.

Our coupled-cluster
computations~\cite{kuemmel1978,bishop1991,bartlett2007,hagen2014}
start from an axially symmetric Hartree-Fock reference state with
prolate deformation~\cite{novario2020,hagen2022}. The inclusion of
full $NNN$ forces increases the computational cost significantly, and
we would therefore like to work with the normal-ordered two-body
approximation~\cite{hagen2007a,roth2012,binder2013}. However, the
normal-ordered two-body Hamiltonian based on a deformed reference
state breaks rotational symmetry. To avoid this problem we follow
\textcite{Frosini:2021tuj} and first perform a spherical Hartree-Fock
computation based on a uniform occupation of the partially filled
shells. The resulting density matrix is then used to make the
normal-ordered two-body approximation, and the Hamiltonian is finally
transformed back to the harmonic oscillator basis. This spherical
two-body Hamiltonian is the starting point for our axially symmetric
Hartree-Fock computation which yields the reference state $\vert
\Phi_0 \rangle$.

Our Hartree-Fock computations use a spherical harmonic oscillator
basis of up to thirteen major shells while the $NNN$ interaction is
further restricted by an energy cut $E_{\mathrm{3max}} = 16
\hbar\Omega$. To gauge the convergence of our results we varied the
harmonic oscillator frequency ($\hbar\omega$) from 10--16~MeV. Due to
computational cost our angular-momentum projected coupled-cluster
calculations are restricted to 8--9 major shells, which are sufficient
to converge quadrupole deformed states in neon
isotopes~\cite{hagen2022}.

We employ the coupled-cluster singles-and-doubles
approximation~\cite{bartlett2007}. For a more accurate
angular-momentum projection than in Ref.~\cite{hagen2022} we use the
bi-variational coupled-cluster energy
functional~\cite{arponen1982,arponen1983,bartlett2007}
\begin{equation}
E^{(J)} = \frac{\langle \widetilde{\Psi} \vert P_J H \vert \Psi \rangle} {\langle \widetilde{\Psi} \vert P_J \vert \Psi \rangle } \ . 
\label{eq:CC_PAV}
\end{equation}
Here $P_J$ is the angular-momentum projection operator (and contains a
rotation operator $R$), $\vert \Psi \rangle \equiv e^T \vert \Phi_0
\rangle$ is the right coupled-cluster state and $\langle
\widetilde{\Psi} \vert \equiv \langle \Phi_0 \vert (1 + \Lambda)e^{-T}
$ is the corresponding left ground-state. In the singles-and-doubles
approximation of coupled cluster, the excitation operator $T$ and the
de-excitation operator $\Lambda$ are truncated at the
two-particle--two-hole ($2p$--$2h$) level. We evaluate
Eq.~(\ref{eq:CC_PAV}) via the disentangled approach by
\textcite{qiu2017}. This approach applies the Thouless
theorem~\cite{thouless1960} to act with the rotation operator $R$ on
the symmetry broken reference state, i.e., $\langle \Phi_0 \vert R =
\langle \Phi_0 \vert R \vert \Phi_0 \rangle \langle \Phi_0 \vert e^V$,
with $V$ being a $1p$--$1h$ de-excitation operator. Next, one expands
$e^V e^T = e^{W_0+W_1+W_2\ldots}$. The series of $np$--$nh$ excitation
operators $W_n$ does not truncate and includes up to $Ap$--$Ah$
excitations for a nucleus with mass number $A$. We only keep the
disentangled amplitudes $W_0, W_1$, and $W_2$ and compute them as the
solution of a set of ordinary differential
equations~\cite{qiu2017}. The truncation at $W_2$ implies that the
projection operator $P_J$ is not treated exactly, and angular momentum
is only approximately a good quantum number. The Supplemental
Material~\cite{supp} might be useful to experts.

\begin{figure*}[!htbp]
    \includegraphics[width=0.99\linewidth]{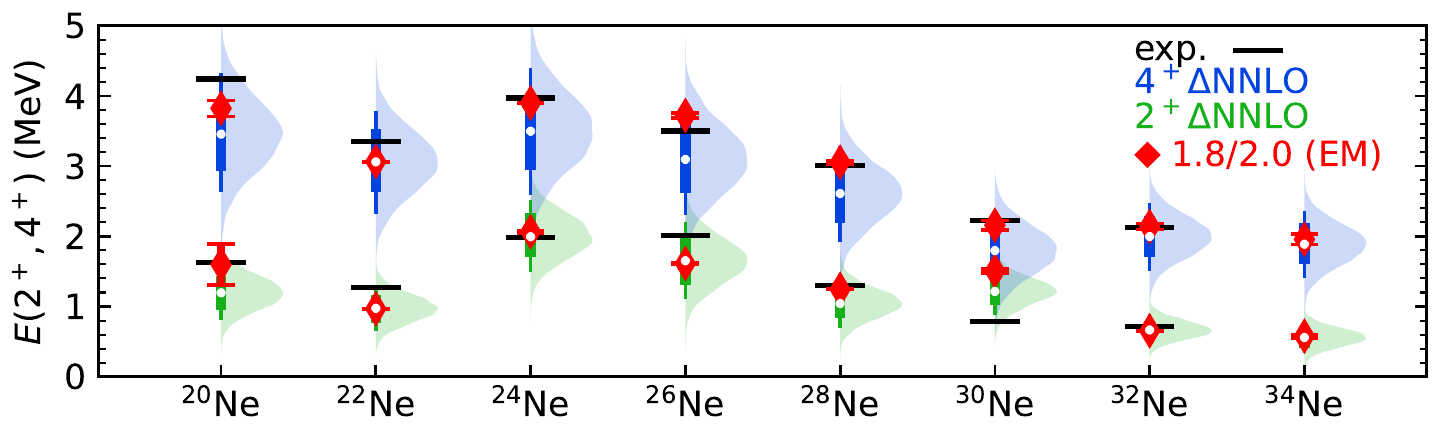}
    \caption{Energies of lowest $2^+$ and $4^+$ states in the even
      nuclei $^{20-34}$Ne, computed using angular-momentum projected
      coupled-cluster with the interaction 1.8/2.0
      (EM)~\cite{hebeler2011} (red diamonds with uncertainties from
      finite model-spaces), and posterior predictive distributions
      from importance resampling using the ensemble of Delta-full
      ($\Delta$) interactions including sampling of method and model
      errors (68\% and 90\% credible intervals as a thick and thin
      vertical bar, respectively, and the median marked as a white
      circle) compared to data (black horizontal bars).}
    \label{fig:neons}
\end{figure*}

\begin{figure}[!htbp]
    \includegraphics[width=0.99\linewidth]{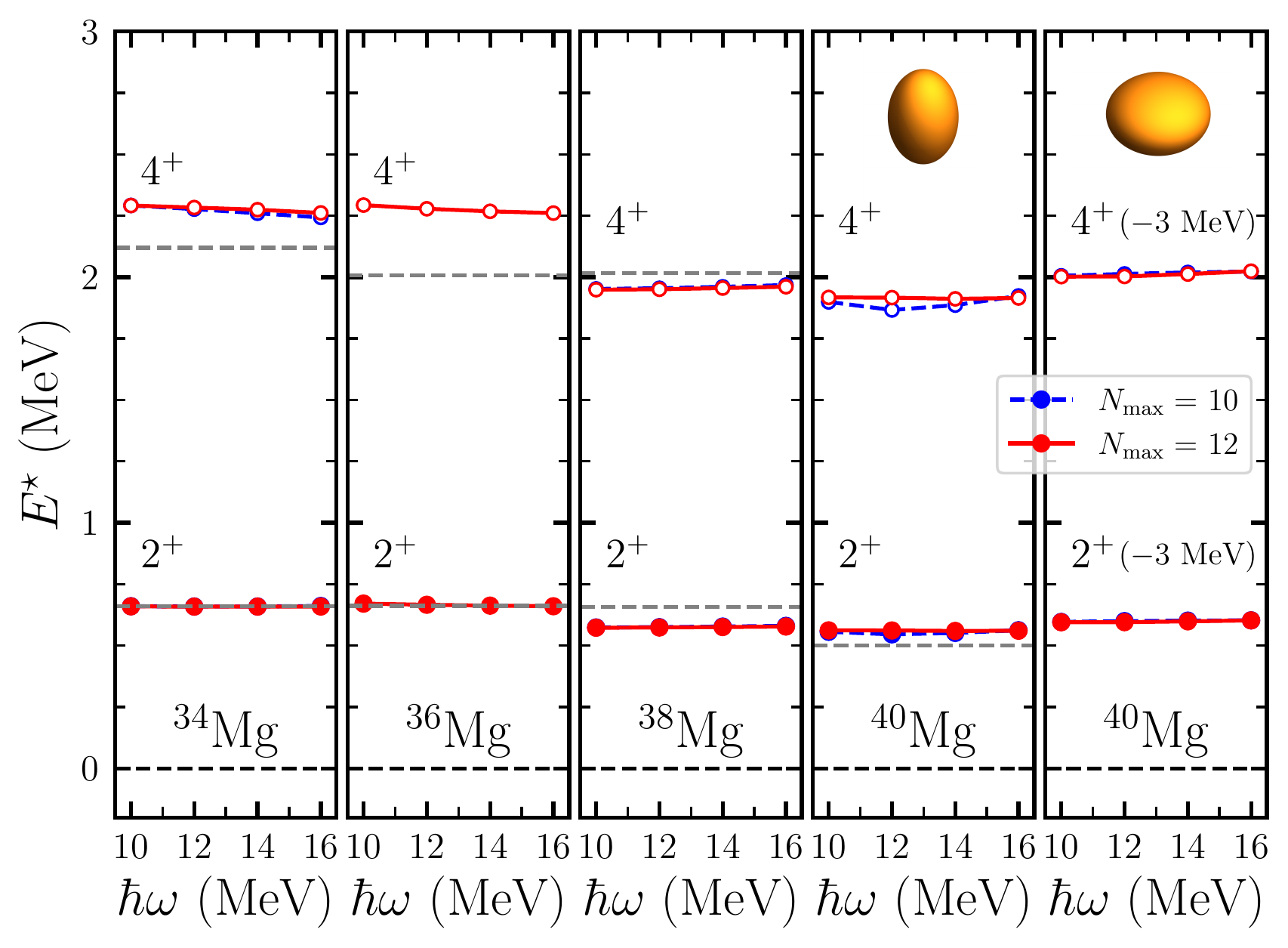}
    \caption{Energies of lowest $2^+$ and $4^+$ states in even nuclei
      $^{34-40}$Mg, as a function of the oscillator frequency and for
      various model spaces, computed using projected Hartree-Fock with
      the chiral interaction 1.8/2.0 (EM)~\cite{hebeler2011} and
      compared to data (dashed horizontal lines). For $^{40}$Mg we
      show both the prolate and oblate rotational bands; the latter
      band head is about 3~MeV above the prolate ground state.}
    \label{fig:magnesiums}
\end{figure}

{\it Results for neon and magnesium isotopes.---}
Figures~\ref{fig:neons} and \ref{fig:magnesiums} show the computed
energies $E(2^+)$ and $E(4^+)$ of the lowest $2^+$ and $4^+$ states in
the even nuclei $^{20-32}$Ne and $^{34-40}$Mg, respectively, and
compares with available data. For neon isotopes, the angular-momentum
projected coupled-cluster results based on the 1.8/2.0 (EM)
interaction include estimates of uncertainties coming from truncated
model-spaces by taking the spread of the results obtained for $N_{\rm
  max} = 6-8$ and $\hbar\omega = 10-16$~MeV.  Using the Delta-full
NNLO interaction ensemble we aim to quantify all relevant
uncertainties. We employ a fixed model-space of $N_{\rm max} = 7$ and
$\hbar\omega = 14$~MeV and assign normally distributed method errors
with relative (one sigma) errors of 10\% (5\%) for $2^+$ ($4^+$)
excitation energies (corresponding to about $100-150$~keV). Similarly,
we assign a 10\% relative EFT truncation error for all excitation
energies.

Overall, theory agrees with data, though the uncertainties are
substantial. For nuclei outside of the island of inversion the
ensemble of NNLO Delta-full interactions give somewhat too compressed
spectra when comparing centroids to data and to the 1.8/2.0 (EM)
interaction (which includes $NN$ forces at N3LO). Our most precise
results are obtained for $^{32,34}$Ne, and the centroids for the
ensemble of Delta-full interactions agree with the results for the
1.8/2.0 (EM) interaction.  For $^{34}$Ne no data is available, and we
predict that the $2^+$ and $4^+$ states are similar to the
corresponding excited states in $^{32}$Ne. The computed $R_{42}\equiv
E(4^+)/E(2^+)$ values~\cite{casten1990} of $^{34}$Ne are $3.37 \pm
0.13$ for the ensemble (68\% confidence interval) and $3.38$ for
1.8/2.0~(EM); both are close to the value $10/3$ of a rigid
rotor. Thus, we expect $^{34}$Ne to be as rotational as
$^{32}$Ne. Overall, \textit{ab initio} theory agrees with data, with
the exception of $E(2^+)$ in $^{30}$Ne which is significantly higher
than experiment and therefore indicates an artificially large $N=20$
shell gap. The persistence of the $N=20$ shell gap in this region of
the nuclear chart was also seen in the projected
generator-coordinate-method computations of
$^{30}$Ne~\cite{Frosini:2021sxj} and coupled-cluster computations of
charge radii of neutron-rich neon and magnesium
isotopes~\cite{novario2020}. These findings hint at deficiencies of
the employed interactions in correctly describing the boundaries of
the island of inversion. One could speculate that our computed $0^+$
ground-state of $^{30}$Ne corresponds to the observed spherical shape
co-existent $0^+$ state in $^{32}$Mg at 1~MeV of excitation
energy~\cite{wimmer2010}.

\begin{figure*}[htp]
    \includegraphics[width=0.99\linewidth]{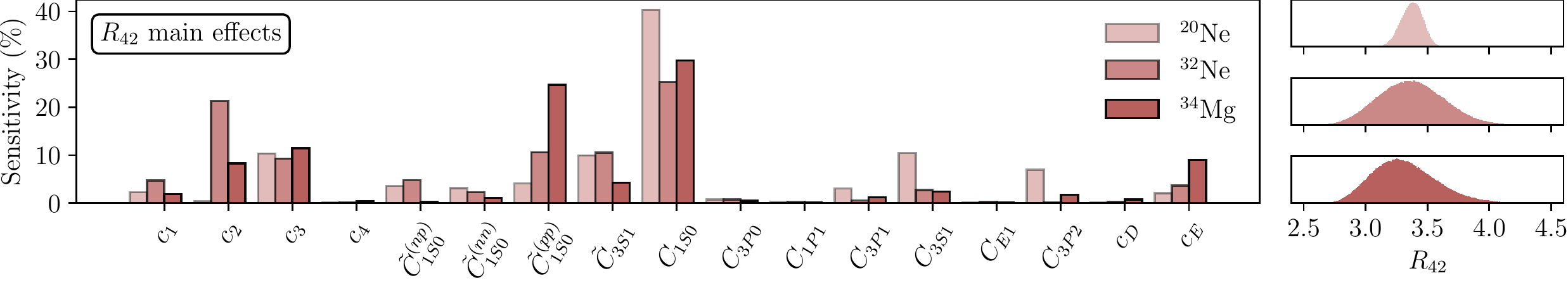}
    \caption{Main effects of the $R_{42}$ deformation measure in
      $^{20,32}$Ne and $^{34}$Mg as obtained in a global
      sensitivity-analysis of $10^6$ emulations of projected
      Hartree-Fock computations using Delta-full chiral EFT at
      NNLO. Left panel: main effects of the low-energy
      constants. Right three panels: Histograms of $R_{42}$ for
      $^{20}$Ne (top), $^{32}$Ne (middle),and $^{34}$Mg (bottom).}
    \label{fig:gsa}
\end{figure*}

For $^{34-40}$Mg we employ the 1.8/2.0 (EM) interaction and use
angular-momentum projected Hartree-Fock in larger model-spaces. This
simplification is justified based on Ref.~\cite{hagen2022} and the
comparison of the rotational bands obtained from Hartree-Fock and
coupled-cluster theory in the neon nuclei. For the dripline nucleus
$^{40}$Mg we also include coupling to the particle continuum by using
a Woods-Saxon basis consisting of bound and scattering states for the
neutron $p_{3/2}$ partial wave, following Ref.~\cite{hagen2016}. The
results are close to data where those are available, see
Fig.~\ref{fig:magnesiums}. One expects an inversion of the $p_{3/2}$
and $f_{7/2}$ single-particle orbitals close to the magnesium
dripline. This is supported by the observation that $^{37}$Mg is a
deformed $p-$wave halo nucleus~\cite{kobayashi2014} and mean-field
computations accounting for deformation and continuum
coupling~\cite{hamamoto2012,hamamoto2016}. Indeed our calculations for
$^{38,40}$Mg show an inversion of the $\Omega^\pi = 7/2^-$ and
$\Omega^\pi = 1/2^-$ single-particle orbitals (where $\Omega$ denotes
the single-particle angular-momentum component along the axial
symmetry axis). We find that $^{34-40}$Mg are all prolate in their
ground-state, with rotational bands that are close to data~(see
Fig.~\ref{fig:magnesiums}). Interestingly, for $^{40}$Mg we also find
an oblate Hartree-Fock state that is close in energy to the prolate
ground state. Performing coupled-cluster calculations for these two
references we find that the oblate band head is about 3~MeV above the
prolate ground-state, indicating an onset of shape co-existence and a
possible interpretation of the third observed excited
state~\cite{crawford2019}. This picture is also consistent with the
Monte-Carlo shell-model computations of
\citet{tsunoda2020}. Figure~\ref{fig:magnesiums} shows both the
prolate and oblate $2^+$ and $4^+$ states, and we observe that the
rotational structure of these two bands are very similar and close to
that of a rigid rotor.

{\it Global sensitivity analysis of deformation.---} We want to
illuminate how the individual terms of the chiral interaction at NNLO
impact deformation in the island-of-inversion nuclei $^{32,34}$Ne and
$^{34}$Mg, and compare this to the deformed and stable nucleus
$^{20}$Ne. To that purpose we perform a variance-based global
sensitivity-analysis~\cite{sobol2001,ekstrom2019} of the ratio
$R_{42}$ in $^{20,32}$Ne and $^{34}$Mg. We partition the total
variance of $R_{42}$ into variances conditional on each of the
low-energy constants in chiral EFT. The dimensionless ratio of a
conditional variance and the total variance in $R_{42}$ is called the
\textit{main effect}, and a greater value indicates a greater
sensitivity of $R_{42}$ to the corresponding low-energy constant.

We consider all 17 low-energy constants of the Delta-full chiral EFT
interaction model in the sensitivity analysis: The leading-order
$S$-wave contacts $\tilde{C}_{^3S_1}$ and $\tilde{C}_{^1S_0}^{(\tau)}$
with $\tau=nn, np, pp$ denoting the isospin projections $1, 0, -1$,
the subleading contacts $C_{^1S_0}$, $C_{^3S_1}$, $C_{^3P_0}$,
$C_{^1P_1}$, $C_{^3P_1}$, and $C_{^3P_02}$ (acting in a partial wave
as indicated by the subscript), and $C_{E_1}$ acting in the the
off-diagonal triplet $S-D$ channel. In addition there are four
subleading pion-nucleon couplings $c_{1,2,3,4}$, as well as the $c_D$
and $c_E$ couplings governing the strengths of the short-range
three-nucleon potential.

The variance integrals underlying the sensitivity analysis are
evaluated on a hypercubic domain centered on the
$\Delta$NNLO$_\text{GO}(394)$
parameterization~\cite{jiang2020}. Drawing on recent Bayesian
analyses~\cite{Wesolowski:2021cni,Svensson:2023twt} we use $\pm
0.05$~GeV$^{-1}$ as the relevant range for each of the pion-nucleon
couplings $c_i$ and $\pm 0.05\times 10^2$~GeV$^{-2}$ for the
sub-leading constants $C_i$. The leading-order contact couplings
$\tilde{C}_i$ are somewhat small (in units of $10^4$~GeV$^{-4}$), in
accordance with naturalness expectations, and their intervals are
limited to $\pm 0.005\times 10^4$~GeV$^{-4}$. We use Monte Carlo
integration to evaluate the variance integrals and this requires
$18\cdot 2^{16} \approx 10^6$ samples to keep the sampling uncertainty
small. Our results are robust when re-scaling all side-lengths of the
hypercube by factors of $1/2$ and $2$. Larger domains result in
noticeable higher-order sensitivities which we did not analyze
further.

It is sufficiently accurate to solve for the excited-state energies
$E(2^+)$ and $E(4^+)$ in $^{20,32}$Ne and $^{34}$Mg using
projection-after-variation
Hartree-Fock~\cite{Frosini:2021sxj,hagen2022}. However, the Monte
Carlo sampling in a global sensitivity analysis requires prohibitively
many projected Hartree-Fock computations. Thus, we develop emulators,
i.e., computationally efficient and accurate models that mimic the
full \textit{ab initio} calculation~\cite{ekstrom2019, Konig:2019adq},
using eigenvector continuation~\cite{frame2018} of angular-momentum
projected Hartree-Fock.
All emulators were trained following a strategy similar to
Ref.~\cite{ekstrom2019}, i.e., using $N_\text{train}=68$ exact
Hartree-Fock states each for the $0^+$, $2^+$, $4^+$ states and
training values for the low-energy constants drawn according to a
space-filling latin hypercube design within 20-30\% variation of of
their $\Delta$NNLO$_\text{GO}(394)$ values. A comparison with 400
exact Hartree-Fock calculations indicates at most $1\%$ discrepancies
(on the 1$\sigma$-level) for the $R_\text{42}$ emulators and even
better for the emulation of excitation energies. The Supplemental
Material~\cite{supp} might be useful to experts.

Figure~\ref{fig:gsa} shows the main effects for $R_{42}$ in
$^{20,32}$Ne and $^{34}$Mg. A majority of the output have $R_{42}
\approx 10/3$ which indicates an axially-deformed rigid rotor and
emergent symmetry-breaking. We find that more than $97\%$ of the
variance in $R_{42}$ is explained via main effects. For all three
nuclei, about $40\%$ of the deformation is driven by the subleading
pion-nucleon coupling $c_3$ and singlet $S$-wave contact. The former
determines the strength of the NNLO two-pion exchange in the
nucleon-nucleon and the three-nucleon potentials. Together with the
main effect of $\tilde{C}$, proportional to the leading singlet
$S$-wave contact, we can explain most of the observed variance in
$R_{42}$. For $^{34}$Mg, with two protons more than $^{20,32}$Ne,
deformation becomes more sensitive to the isospin-breaking $S$-wave
contact in the proton-proton channel. Towards the neutron dripline,
deformation is more sensitive to the short-range three-nucleon forces
and the pion-nucleon coupling $c_2$.

We can also use the ensemble employed in Fig.~\ref{fig:neons} to probe
what impacts deformation. The relevant parts of the nuclear
interaction can be identified by studying correlations between the
observable $R_{42}$ in $^{32}$Ne and individual low-energy
constants. We find that the correlation is strongest for $C_{^1S_0}$
(with a correlation coefficient $r=0.73$, i.e., an increase in the
repulsive $C_{^1S_0}$ increases deformation), but it is also sizeable
for the three-nucleon contact $c_E$. Comparing these results with the
conditional variances from the global sensitivity analysis confirms
the importance of pairing via the $^1S_0$ channel. We note that the
domain of low-energy constants used in the global sensitivity analysis
is smaller than the non-implausible volume spanned by the interaction
ensemble from history matching.

{\it Conclusions.---} We reported on {\it ab initio} computations of
neutron-rich neon and magnesium nuclei. For neon, an ensemble of
chiral interactions and the 1.8/2.0 (EM) interaction yield accurate
results (except for $N=20$), and we predict that $^{34}$Ne is well
deformed. For $^{34-38}$Mg our calculations with the 1.8/2.0 (EM)
interaction are close to data, and we predict a prolate ground-state
rotational band and an excited oblate band in $^{40}$Mg. A global
sensitivity analysis and a study of correlations reveal that a few
low-energy constants strongly impact deformation. This is the first
step in understanding nuclear deformation at a high resolution scale.

\begin{acknowledgments}
This work was supported by the U.S. Department of Energy, Office of
Science, Office of Nuclear Physics, under Award Nos.~DE-FG02-96ER40963
and DE-SC0018223, by SciDAC-5 (NUCLEI collaboration), by the Quantum
Science Center, a National Quantum Information Science Research Center
of the U.S. Department of Energy, by the European Research Council
(ERC) under the European Unions Horizon 2020 research and innovation
program (Grant Agreement No. 758027), the Swedish Research Council
(Grants No. 2017-04234, No. 2020-05127 and No. 2021-04507). Computer
time was provided by the Innovative and Novel Computational Impact on
Theory and Experiment (INCITE) programme. This research used resources
of the Oak Ridge Leadership Computing Facility located at Oak Ridge
National Laboratory, which is supported by the Office of Science of
the Department of Energy under contract No. DE-AC05-00OR22725 and
resources provided by the Swedish National Infrastructure for
Computing (SNIC) at Chalmers Centre for Computational Science and
Engineering (C3SE), and the National Supercomputer Centre (NSC)
partially funded by the Swedish Research Council through grant
agreement no.\ 2018-05973.
\end{acknowledgments}

\bibliography{arxiv}

\clearpage
%\newpage

% \appendix

\section{Supplemental Material: How chiral forces shape neutron-rich Ne and Mg nuclei}

\subsection{Convergence of projection-after-variation Hartree-Fock and the disentangled coupled-cluster approach} 

\begin{figure}[!htbp]
    \includegraphics[width=0.99\linewidth]{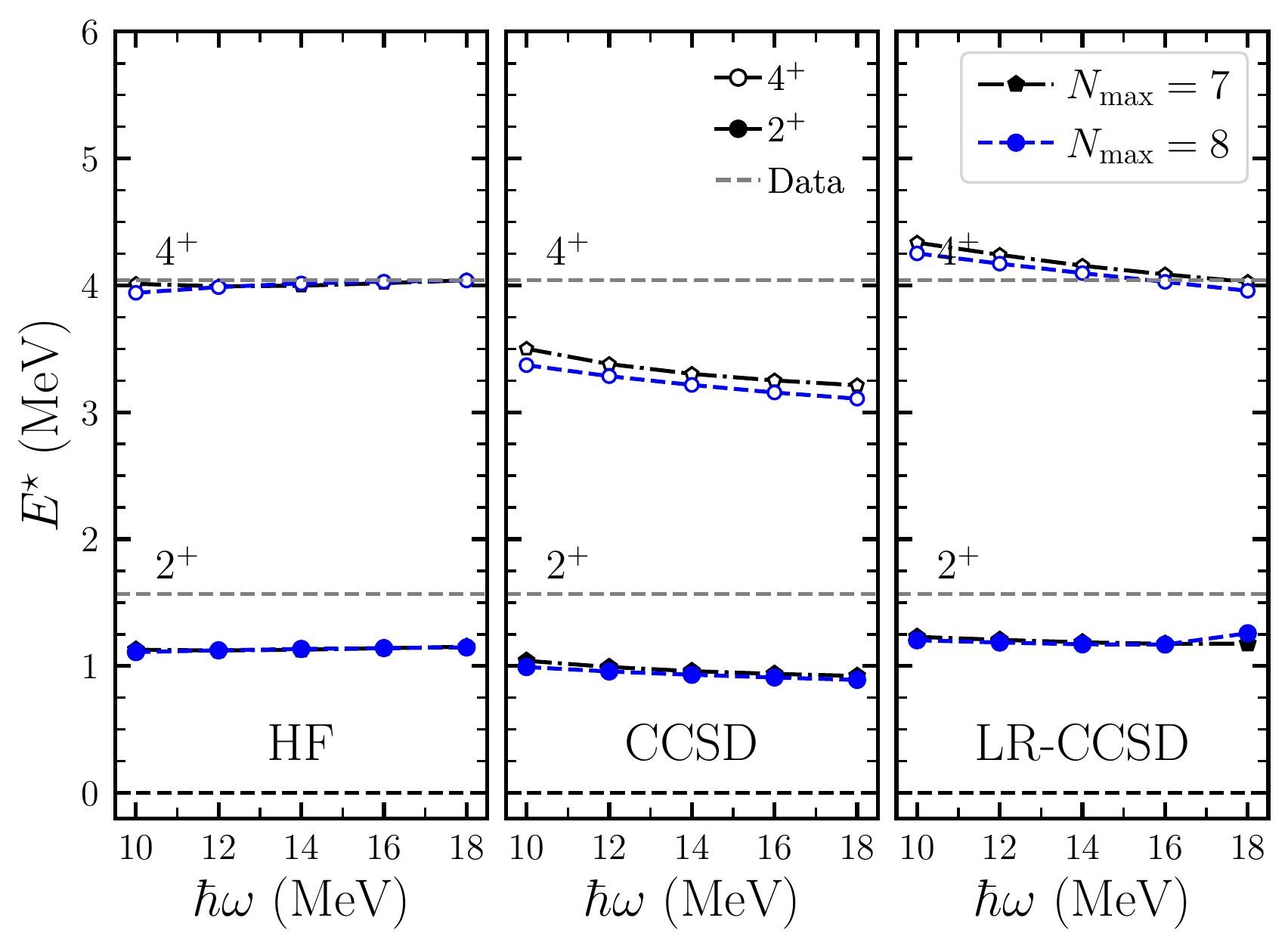}
    \caption{Comparison between projection-after-variation Hartree
      Fock (HF), projected coupled-cluster (CCSD), and projected
      linear-response coupled-cluster method (LR-CCSD) for the excited
      $2^+$ and $4^+$ in $^{20}$Ne using the NNLO$_{\mathrm{opt}}$
      nucleon-nucleon interaction. The dashed lines show the results
      from symmetry-adapted no-core shell-model computations using the
      same interaction.}
    \label{fig:ne20_benchmark}
\end{figure}

Figure ~\ref{fig:ne20_benchmark} shows the $2^+$ and $4^+$ rotational
states in $^{20}$Ne obtained with the NNLO$_{\mathrm{opt}}$
nucleon-nucleon interaction~\citeapp{ekstrom2013}, and using the
projection-after-variation Hartree-Fock and disentangled approaches
with the ``naive'' and the bi-variational (linear-response)
coupled-cluster ansatz~\cite{qiu2017}, respectively. We observe that
both states are well converged with respect to the model-space size
for projection-after-variation Hartree-Fock. For the disentangled
approach there is a slightly larger dependence on the employed
model-space. This is likely so because the disentangled
coupled-cluster approach does not restore the broken symmetry
exactly~\cite{qiu2017,hagen2022}. We also observe that
projection-after-variation Hartree-Fock and the bi-variational
disentangled coupled-cluster agree with benchmarks from
symmetry-adapted no-core shell-model
calculations~\citeapp{heller2022new}, while the ``naive'' disentangled
approach yields more compressed spectra.

\begin{figure*}[!htbp]
    \includegraphics[width=0.99\linewidth]{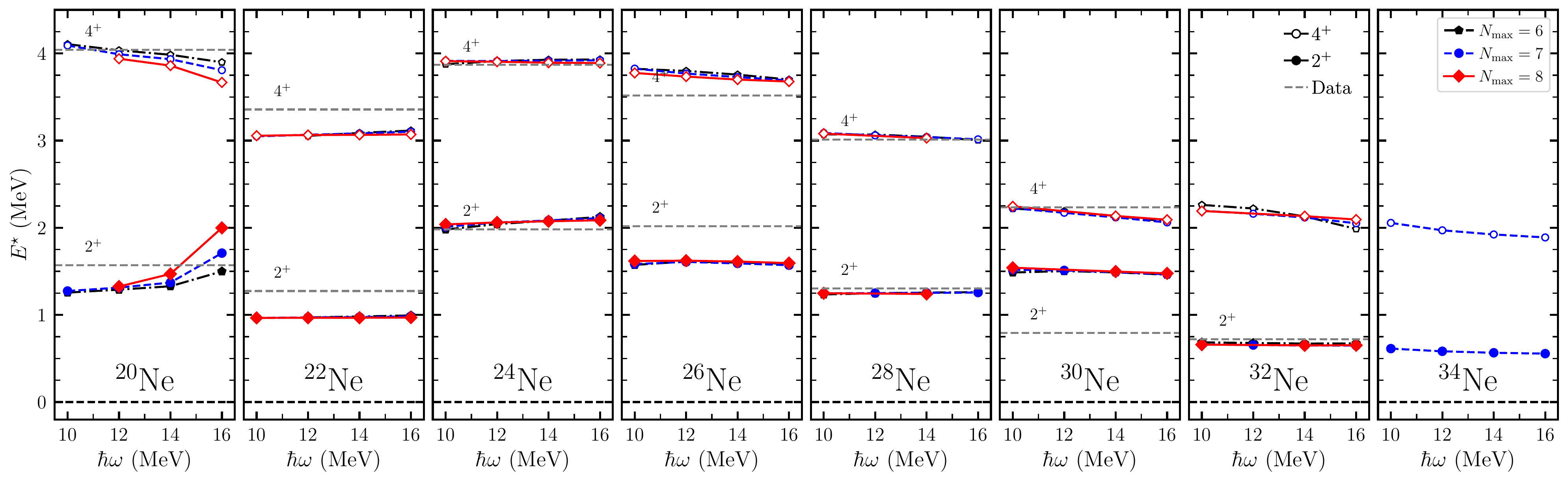}
    \caption{Energies $E$ of the lowest $2^+$ and $4^+$ states in even
      nuclei $^{20-34}$Ne, computed with the interaction 1.8/2.0 (EM)
      from chiral effective field theory~\cite{hebeler2011} and
      compared to data.}
    \label{fig:neons-magic}
\end{figure*}
Figure~\ref{fig:neons-magic} shows the $2^+$ and $4^+$ excited states
in $^{20-34}$Ne obtained with the interaction 1.8/2.0
(EM)~\cite{hebeler2011} and compared to data for different model-space
sizes. We find (with the exception of $^{20}$Ne at
$\hbar\omega=16$~MeV) that the states are well converged with respect
to model-space size. The variation of the results with respect to the
model-space is shown as an uncertainty in Fig.~\ref{fig:neons} in the
main text.

\subsection{Emulating projection-after-variation
Hartree-Fock using eigenvector continuation}  

To construct the Hartree-Fock emulators we exploit that the Delta-full
NNLO Hamiltonian can be written in a form that depends linearly on the
17 low-energy constants $(\vec{\alpha})$ of interest, i.e.,
\begin{equation}
H(\vec{\alpha}) = H_{0} + \sum_{i=1}^{17}  \alpha_i H_i.
\label{eq:full_hamiltonian}
\end{equation}
Here, $H_0=T_{\rm{kin}}+V_0$ is independent of the low-energy
constants we consider here and includes the intrinsic kinetic energy
$T_{\rm{kin}}$ and potential terms $V_0$ such as one-pion exchange,
leading two-pion exchange, and the
Fujita-Miyazawa~\citeapp{fujita1957} three-nucleon interaction.

To proceed we write the full Hamiltonian in normal-ordered form with
respect to an arbitrary Hartree-Fock reference state, i.e. $H =
E_{\mathrm{HF}} + F + W$. Here $E_{\mathrm{HF}}$ corresponds to the
Hartree-Fock (vacuum) energy, $F$ is the Fock-matrix (i.e. the
normal-ordered one-body term), and $W$ contains two- and three-body
normal-ordered terms. We note that $F$ is diagonal in the Hartree-Fock
representation. We can then write the one-body Hartree-Fock
Hamiltonian as $H_{\rm{HF}} = E_{\rm{HF}} + F$.  We seek to emulate
the Hartree-Fock energy for some target value $\vec{\alpha} =
\vec{\alpha}_{\circledcirc}$.  To that end, we expand the target
wave-function in a set of Hartree-Fock states $\vert
\phi_{\circledcirc} \rangle = \sum_i^{n_t} c_i \vert \phi_i\rangle$,
where $|\phi_i\rangle$ is a Hartree-Fock state obtained for the
training point of low-energy constants $\vec{\alpha}_i$, i.e.
\begin{equation}
    H_{\rm{HF}}(\vec{\alpha}_i)|\phi_i\rangle =
    E_{\rm{HF}}(\vec{\alpha}_i)|\phi_i\rangle\ ,
\end{equation}
and $H_{\rm{HF}}(\vec{\alpha}_i)$ is the Hartree-Fock Hamiltonian for
$\vec{\alpha}_i$. The emulator step relies on expressing the
eigenstates of the target Hamiltonian
$H_{\rm{HF}}(\vec{\alpha}_{\circledcirc})$ as the solution to the
generalized eigenvalue problem
\begin{equation} 
\label{eq:general_eigenvalue}
\sum_{ij}\langle {\phi}_i \vert
H_{\rm{HF}}(\vec{\alpha}_{\circledcirc}) \vert \phi_j \rangle c_j =
E_{\circledcirc} \sum_{ij} \langle {\phi}_i \vert \phi_j \rangle c_j\ ,
\end{equation}
where $E_{\circledcirc} \approx
E_\text{HF}(\vec{\alpha}_\circledcirc)$ and the sums run over the
$N_\text{train}$ eigenstates obtained during the training step. In the
following we write $h = H_{\rm{HF}}$ for simplicity of notation. To
evaluate the norm and Hamiltonian kernels we utilize the Thouless
theorem~\cite{thouless1960}, and follow an approach similar to the one
used for the disentangled coupled-cluster approach~\cite{qiu2017}.

The Hartree-Fock solution $\vert {\phi}_i \rangle $ for training point
$\vec{\alpha}_i$ is connected to the Harmonic oscillator unrestricted
reference state $\vert \Phi_0 \rangle$ via a unitary transformation,
i.e. $\vert \phi_i \rangle = U_i \vert \Phi_0 \rangle$. The norm and
Hamiltonian kernels for eigenvector continuation are
\begin{eqnarray}
\label{eq:norm}
 \langle {\phi}_i \vert \phi_j \rangle & = & \langle \Phi_0 \vert
 \mathcal{O}_{ij} \vert \Phi_0\rangle, \\
 \label{eq:hamiltonian}
   \langle {\phi}_i \vert h(\vec{\alpha}_{\circledcirc}) \vert \phi_j
   \rangle & = & \langle \Phi_0 \vert \mathcal{O}_{ij}
   h_j(\vec{\alpha}_{\circledcirc}) \vert \Phi_0 \rangle.
\end{eqnarray}
Here $\mathcal{O}_{ij} = U_i^\dagger U_j $ is a unitary matrix, and
$h_j(\vec{\alpha}_{\circledcirc})$ is the target Hamiltonian
normal-ordered with respect to the Hartree-Fock state of training
point $\vec{\alpha}_j$, i.e. $h_j(\vec{\alpha}_{\circledcirc}) =
u_j^\dagger h(\vec{\alpha}_{\circledcirc}) u_j.$ We note that the
Hamiltonian kernel in Eq.~(\ref{eq:hamiltonian}) is not symmetric, and
the Fock matrix, $F_j(\vec{\alpha}_{\circledcirc})$, is not diagonal
for $\vec{\alpha}_{\circledcirc} \neq \vec{\alpha}_{j}$. This is an
important point and it ensures that -- at a training point -- the
solution of the generalized eigenvalue problem is indeed an eigenstate
of the Hartree-Fock Hamiltonian (and not a generator-coordinate-method
solution that is lower in energy).  The norm kernel for non-orthogonal
reference states is given by~\citeapp{lowdin1955}
\begin{equation}
\langle \Phi_0 \vert \mathcal{O}_{ij} \vert \Phi_0\rangle =
\mathop{\rm det}\left({\mathcal{O}_{ij}^{hh}}\right) \ ,
\end{equation}
here $\mathcal{O}_{ij}^{hh}$ is the matrix of overlaps between
occupied (hole) states in $\langle \phi_i\vert $ and $\vert
\phi_j\rangle$. To evaluate the Hamiltonian kernel we utilize the
Thouless theorem~\cite{thouless1960} and write,
\begin{equation}
\label{eq:thouless}
\langle \Phi_0 \vert \mathcal{O}_{ij} = \langle \Phi_0 \vert
\mathcal{O}_{ij} \vert \Phi_0 \rangle \langle \Phi_0 \vert e^V,
\end{equation}
with $V$ being a $1p$--$1h$ de-excitation operator. The matrix
elements of $V$ in the hole-particle ($hp$) space is given by the
matrix product~\cite{qiu2017},
\begin{equation}
V^{hp} = \left( \mathcal{O}_{ij}^{hh}\right)^{-1} \mathcal{O}_{ij}^{hp}.
\end{equation}
Inserting Eq.~(\ref{eq:thouless}) into Eq.~(\ref{eq:hamiltonian}) we
obtain the algebraic equation,
\begin{equation}
 \langle {\phi}_i \vert H_{\rm HF}(\vec{\alpha}_{\circledcirc}) \vert
 \phi_j \rangle = \langle {\phi}_i \vert \phi_j \rangle \left( E_{\rm
   HF} + \sum_{hp} V^h_p F_h^p \right) \ .
\end{equation}
Here $E_{\rm HF}$ and $F$ are the vacuum energy and one-body normal
ordered terms, respectively, of $H_{\rm
  HF}(\vec{\alpha}_{\circledcirc})$ with respect to $\vert \phi_j
\rangle $.

We note that the norm and Hamiltonian Hartree-Fock kernels can also be
evaluated using a generalized Wick's
theorem~\citeapp{lowdin1955,hoyos2012, robledo2020,burton2021}. We
verified that this alternative approach gives results that agree with
the one used in this work. Having obtained the emulated target
Hartree-Fock state by diagonalizing the generalized non-symmetric
eigenvalue problem in Eq.~(\ref{eq:general_eigenvalue}), we evaluate
the projected target Hartree-Fock energies from
\begin{equation}
E^{(J)}_{\circledcirc} = \frac{\langle \phi_{\circledcirc} \vert P_J
  H(\vec{\alpha}_{\circledcirc}) \vert \phi_{\circledcirc} \rangle}
{\langle \phi_{\circledcirc} \vert P_J \vert \phi_{\circledcirc}
  \rangle }\ .
\end{equation}
Here the full target Hamiltonian in Eq.~(\ref{eq:full_hamiltonian})
enters, and $P_J$ is the projection operator.

\subsection{Sensitivity analysis of deformation in neutron-rich neon isotopes} 
\begin{figure*}[!htb]
    \includegraphics[width=0.9\linewidth]{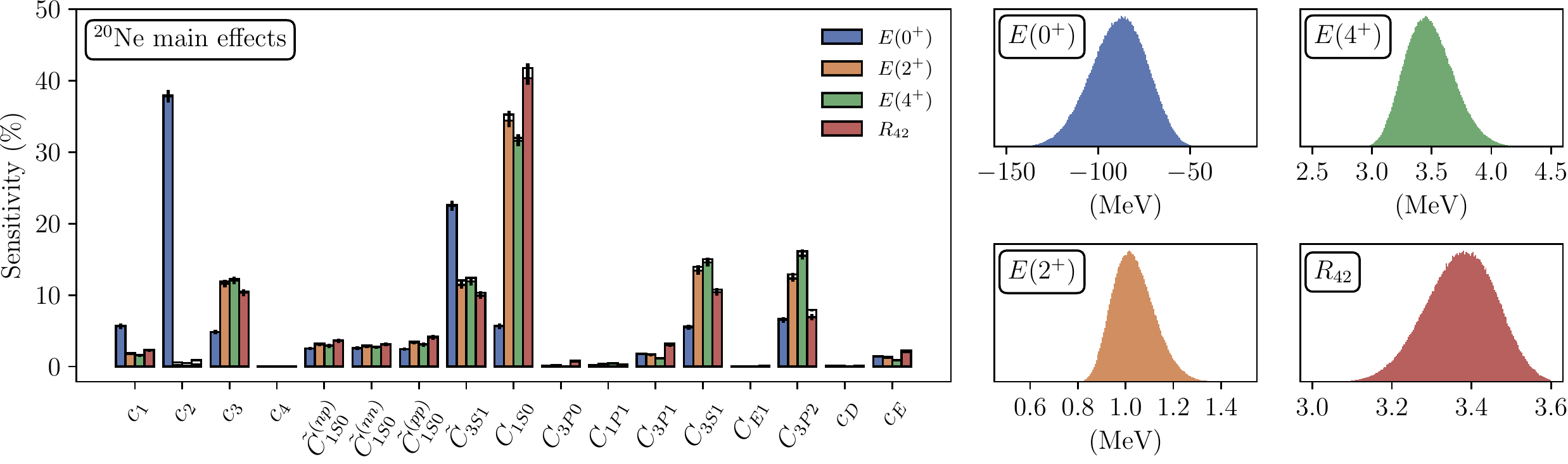}
    \includegraphics[width=0.9\linewidth]{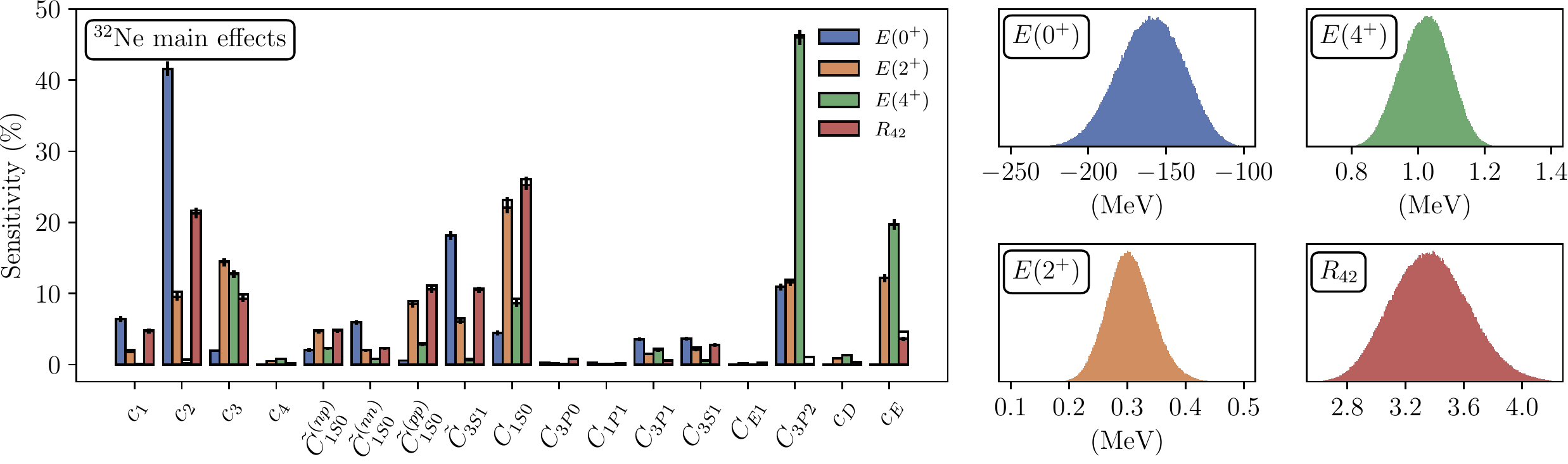}
    \includegraphics[width=0.9\linewidth]{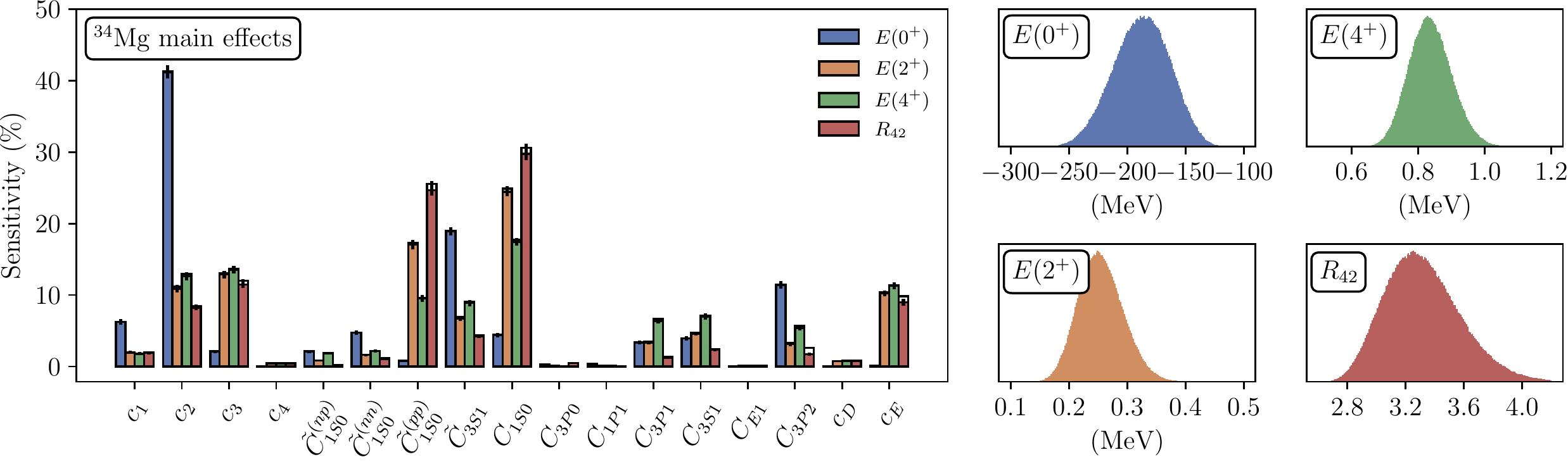}
    \caption{Results of the global sensitivity-analysis for the
      structure of $^{20,32}$Ne and $^{34}$Mg (top, center, and bottom
      panels). All Monte Carlo samples obtained as in the main
      text. Left panels: the mains effects (colored bars) for $R_{42}$
      ground state energy ($E_{gs}$) and excited $2^+$ and $4^+$
      states. The vertical black bars indicate the respective 95\%
      confidence intervals of the sensitivity indices as obtained
      using bootstrapped sampling. Here we also include total effects
      (white bars on top). Groups of right four panels: histograms
      displaying the variation of the 1,179,648 samples for each
      output. All energies were obtained using an emulator based on
      eigenvector continuation of projected Hartree-Fock.}
    \label{fig:all_gsa}
\end{figure*}

Figure~\ref{fig:all_gsa} shows the main effects for the $R_{42}$
ratios and the ground and excited state energies (i.e., $E(0^+)$,
$E(2^+)$, and $E(4^+)$) in $^{20,32}$Ne and $^{34}$Mg.  All results
are based on the same set of samples as in the main text. Here, we
also include the total effects~\citeapp{SALTELLI2010259} (shown as
white bars on top of colored bars of the main effects) and they are
nearly identical to the main effects. This indicates that the (sum of)
higher-order sensitivities are very small in the present domain.

The main effects for the $R_{42}$ ratios is discussed in the main
text. Here we comment of the energies $E(0^+)$, $E(2^+)$, and
$E(4^+)$. There are three main trends in
Fig.~\ref{fig:all_gsa}. First, the variance of $E(0^+)$, i.e., the
energy of the ground state, is explained almost entirely by the
subleading pion-nucleon coupling $c_2$ and the leading $S-$wave
contact $\tilde{C}_{^3S_1}$ in all three nuclei. The latter coupling
is directly proportional to the deuteron binding energy. Second, the
main effects are distributed over more couplings for the nuclei
$^{32}$Ne and $^{34}$Mg that are closer to the neutron dripline.
Third, the ground and excited state energies exhibit different main
effect patterns for neutron-rich Ne and Mg compared to $^{20}$Ne
indicating that the structure of their respective wave functions
likely differ.

\begin{figure*}[!htbp]
    \includegraphics[width=0.32\linewidth]{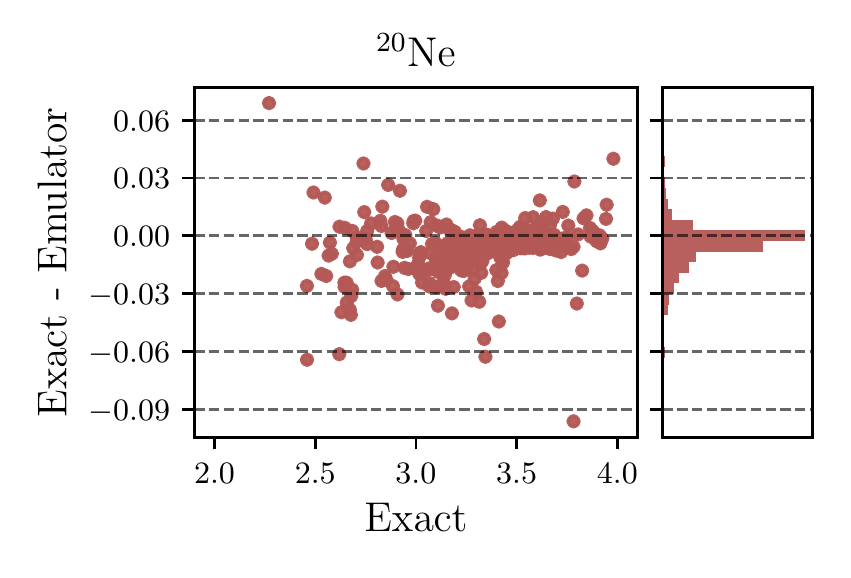}
    \includegraphics[width=0.32\linewidth]{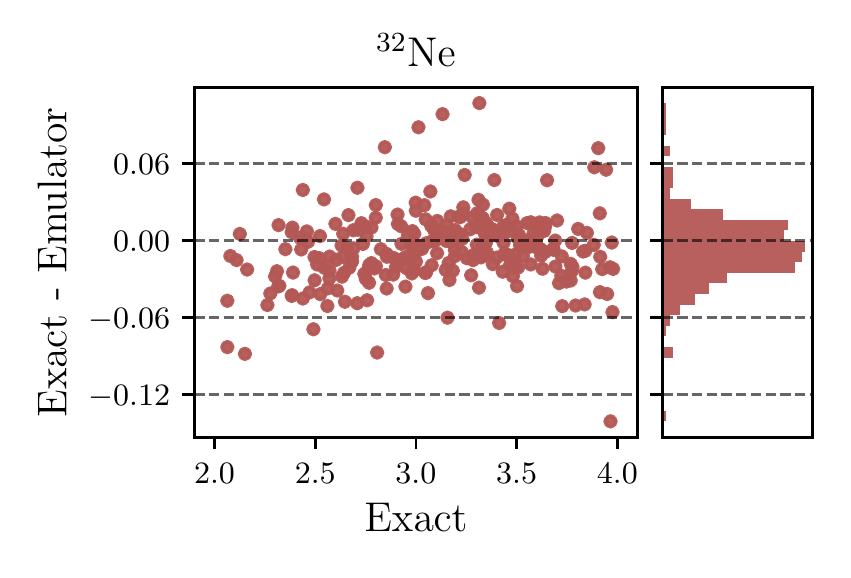}
    \includegraphics[width=0.32\linewidth]{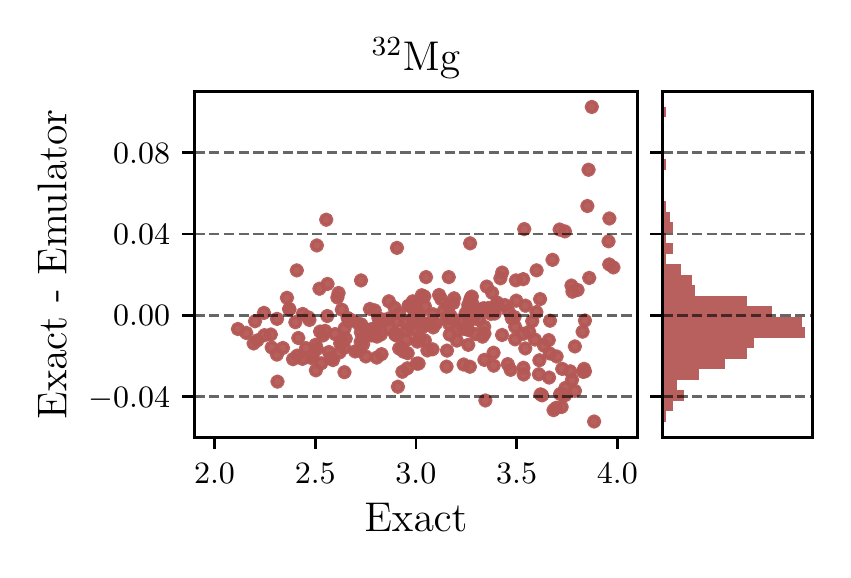}
    \caption{Accuracy of the $R_{42}$ emulator as measured by the
      difference with respect to 400 exact calculations. Shown are the
      subsets of results for which $R_{42}\in[2,4]$, i.e., the range
      relevant to the global sensitivity analysis. The horizontal
      dashed lines indicate precision as they are spaced by two
      standard deviations of the projected histograms (shown in side
      panels).}
    \label{fig:xval_gsa}
\end{figure*}

Figure~\ref{fig:xval_gsa} shows the accuracy of the $R_{42}$ emulators
as measured by comparison with exact projected Hartree-Fock
calculations. The standard deviation of the differences indicates a
relative precision of $1\%$ for the interval of $R_{42}$ values
relevant to the global sensitivity analysis. The emulation of
excitation energies is accurate on the 10 keV level.

\subsection{Correlation analysis of deformation in neutron-rich neon isotopes} 

\begin{center}
\begin{figure}[hb!]
    \includegraphics[width=0.99\textwidth]{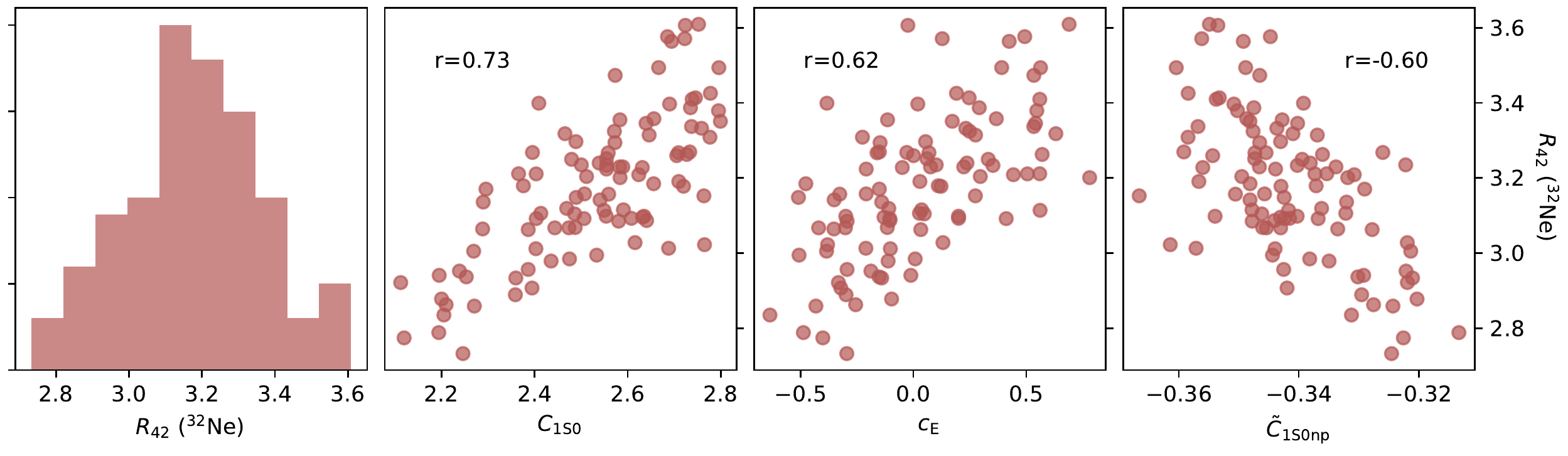}
    \includegraphics[width=0.99\textwidth]{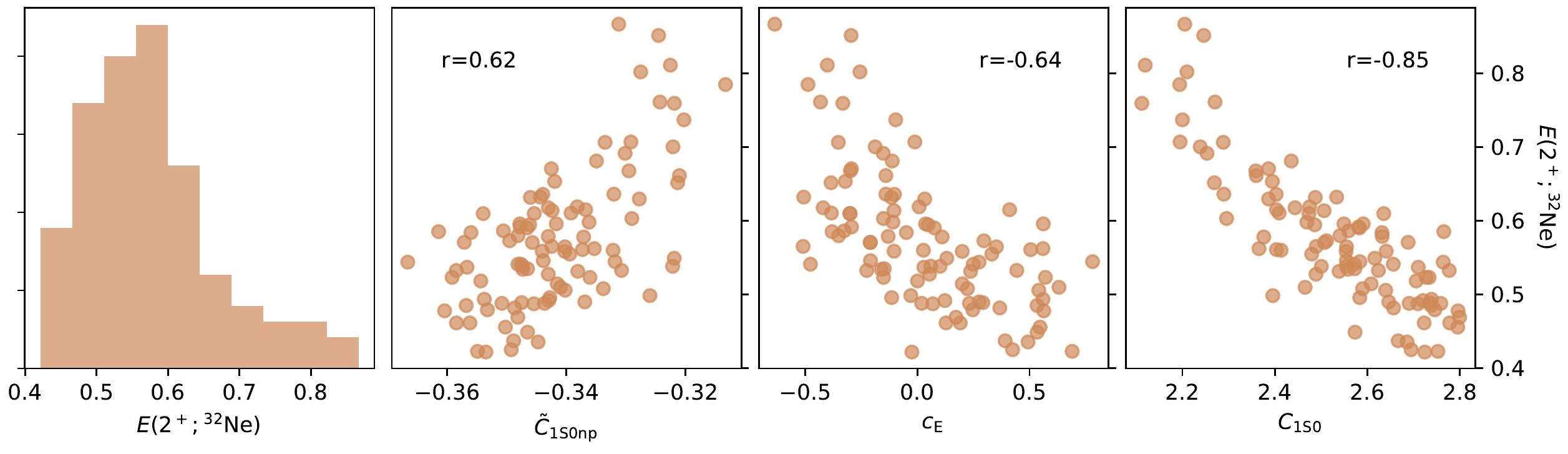}
    \caption{Correlation between the low-energy constants of the
      Delta-full NNLO Hamiltonian and the observables $R_{42}$ (upper
      row), $E(2+)$ (lower row) of $^{32}$Ne. The scatter plots
      include the $\approx 100$ interaction samples from the history
      matching ensemble while the histograms show the distribution for
      the respective observable. The Pearson correlation coefficient,
      $r$, is extracted for the samples for each pair of
      variables. All other low-energy constants have correlation
      coefficients with an absolute magnitude smaller than 0.4 and are
      not shown.}
    \label{fig:ne20-R42_pars}
\end{figure}
\end{center}

Figure~\ref{fig:ne20-R42_pars} shows the strongest correlations
between the low-energy constants of the Delta-full NNLO Hamiltonian
and the observables $R_{42}$ and $E(2+)$ of $^{32}$Ne for the ensemble
of interactions from history matching.
\onecolumngrid
\twocolumngrid

\clearpage
%\bibliographystyleapp{apsrev4-1}
\bibliographystyleapp{plain}
\bibliographyapp{app}

\end{document}